\documentclass[12pt]{iopart}
\usepackage{epsfig}
\usepackage{rotating}

\newcommand{\cph}{\ensuremath\mathrm{cycles}/\sqrt{\mbox{Hz}}}
\newcommand{\mcph}{\ensuremath\mathrm{{\mu}cycles}/\sqrt{\mbox{Hz}}}

\newenvironment{idlplot}[2]{
\begin{figure}[#1]\centering
\includegraphics[width=3.5in, height=6in, angle=90]{#2}
}{\end{figure}}

\begin{document}


\article[A Demonstration of LISA Laser Communication]{}
{A Demonstration of LISA Laser Communication}

\author{S E Pollack\dag\ and R T Stebbins\ddag}
\address{\dag\ JILA, University of Colorado, Boulder, CO  80309-0440\\
    {\it present address:} CENPA, Nuclear Physics Laboratory, 
    University of Washington, Seattle, WA  98195-4290}
\address{\ddag\ NASA/GSFC Code 661, Greenbelt, MD  20771}
\ead{scott.pollack@colorado.edu}

\begin{abstract}
Over the past few years questions have been raised concerning
the use of laser communications links between sciencecraft
to transmit phase information crucial to the reduction
of laser frequency noise in the LISA science measurement.
The concern is that applying medium frequency phase
modulations to the laser carrier could
compromise the phase stability of the LISA fringe signal.
We have modified the table-top interferometer presented
in \cite{ifo} by
applying phase modulations to the laser beams in order
to evaluate the effects of such modulations on the 
LISA science fringe signal.  We have demonstrated
that the phase resolution of the science signal
is not degraded by the presence of
medium frequency phase modulations.
\end{abstract}


\submitto{\CQG}


\section{Introduction\label{intro}}

There are numerous interferometric fringe signals in the LISA interferometer.
Most of these fringes are created by heterodyning a local laser on one
spacecraft with a remote laser located on another spacecraft.
The characteristics of the received beam cause the
LISA fringe to have several unique aspects.
In \cite{ifo} we presented a test-bed interferometer designed
to produce LISA-like fringes.  
We captured three aspects of the LISA-like fringe: 
(1) a baseband fringe frequency which ranges from zero to tens of MHz, 
(2) the fringe sweeps at a rate up to 1\,Hz/s, and
(3) the fringe has a small signal level 
resulting from the heterodyning of 100\,pW and 0.5\,mW laser beams.  
We mentioned in \cite{ifo} the utility of our setup to
demonstrate and investigate other aspects of the LISA
interferometry measurement system.  In this paper
we discuss a plan for laser communication and
demonstrate this plan with our table-top interferometer.

The three LISA sciencecraft operating in concert
constitute the gravitational wave detector.
The photoreceivers on each optical bench 
in the LISA constellation record phase information.  
The measurement
of gravitational waves requires at least two working arms
of the LISA constellation.  In addition, the lasers on-board
each sciencecraft must be related to each other somehow in
order to make the proper phase comparisons which lead
to a strain measurement.  Each laser is independently
stabilized to a frequency reference or to the LISA arms
\cite{arm1, arm2, arm3}.
An alternative stabilization scheme is to stabilize all lasers
to one master laser, and yet another is to allow the six lasers
to run independently (see e.g., \cite{TSSA}).
Regardless of the choice to stabilize, or which
stabilization method will be implemented,
the frequency difference
and noise between the lasers must be measured and canceled out
to reach the required strain sensitivity for LISA.  
The frequency noise correction system is based on an algorithm known
as time-delay interferometry (TDI)
which is well discussed in the literature
(see e.g., \cite{TEA} and references therein).

\section{LISA Telemetry and Ranging}

Time-delay interferometry (TDI) is the process
of combining phase information with time-delays inserted
to cancel out phase noise.  In particular, the phase noise
due to frequency jitter in the lasers can be reduced dramatically.
TDI requires not only the phase information produced
by the interferometric fringe of LISA, but also some knowledge
of the relative jitter between the ultra-stable oscillators (USOs)
onboard each sciencecraft.

\subsection{Ultra-Stable Clock Oscillator\label{s:uso}}

Onboard each LISA sciencecraft is an ultra-stable
oscillator (USO) which provides the reference frequency
for the LISA frequency distribution system (FDS).  The FDS
provides frequencies to several key LISA systems, including
the phase measurement system,
the laser stabilization system, 
the frequency noise correction system,
and the telemetry system.

Jitter in the USOs on each sciencecraft
will appear as jitter in the phase of the LISA fringe signal.
Therefore it is imperative to actively measure the phase noise 
of the USOs onboard LISA.  

In each optical assembly
a high frequency sideband derived from the USO will
be modulated onto the laser beam.  This will be 
referred to throughout this paper interchangeably
as the USO sideband, the USO subcarrier, or just simply the subcarrier.
The subcarrier will be present on each laser beam
so that it will be sent both ways on each of the arms and
between the optical benches contained in each sciencecraft.
Beating of this sideband with the sideband produced by the
local USO produces the error signal used for frequency
noise correction.

The current baseline plan for modulating the USO onto the
laser beam is as follows.
Each laser will consist of a
neodymium yttrium-aluminum-garnet (Nd:YAG)
master oscillator
fiber-coupled to a ytterbium (Yb:YAG) fiber amplifier.
The current plan is to insert a fiber modulator before
the fiber amplifier.
The fiber modulator will be an electro-optic (EO) device
used for phase modulation.
The modulation depth of the EO will be about 10\%
to facilitate transfer of this signal.

Naturally there is some requirement on the measurement
of the USO sideband.  This is determined by the
requirement on sampling synchronization from TDI.
The phase measurement requirement on the USO sideband
is $10^{-4}\,\cph$ at 1~mHz \cite{imsreq}.

The frequency selection of the USO sideband has two constraints:
(1) it must be of high enough frequency to minimize 
the effect of shot noise,
and (2) it must be of low enough frequency to be
realized with an EO modulator.
The current baseline design for LISA is to utilize sideband-sideband
beatnotes to allow the
USO sideband frequency to be much higher
than in a sideband-carrier scheme.
This further reduces the
effect of shot noise in the phase measurement of the USO.  The
USO sideband frequency currently is set at 2\,GHz \cite{refarch}.
The USO sidebands on each sciencecraft will be slightly different.
For instance they might be 2\,MHz apart.
In this way the sideband-sideband beatnote will be $\pm 2\,\mathrm{MHz}$
around the carrier-carrier beatnote between the incoming and local
laser beams.  
The carrier-carrier beatnote contains the science information (i.e.,
gravitational wave signals).  The frequency of all the beatnotes
will be shifted by the Doppler frequency due to orbital
motions of the sciencecraft over the course
of a year, roughly $\pm$ 20~MHz.

\subsection{Ranging Tones}

In addition to the USO signal on the LISA laser beam,
ranging tones may be modulated onto the beam.
The TDI algorithm for laser frequency noise
correction requires absolute knowledge of
the armlength differences of the LISA constellation to
tens of meters accuracy \cite{TEA}.
This requirement is not easily met using ground
tracking alone.  Combining the ground tracking
with a laser-based ranging measurement will yield
an estimate of the armlengths to the required accuracy.
One alternative to using ranging tones would be to 
determine the distance using the new technique of 
time-delay interferometric ranging whereby the ranging 
distances are solved for by a minimization of the noise in the
TDI variables rather than actually being measured \cite{TDIR}.
However, for now we will assume that ranging tones will still be
required for LISA.

To prevent contamination of the LISA science data,
the ranging tones will be phase modulated onto the
USO signal before phase modulation onto the laser beam.
This is the source of the terminology of the USO sideband
being referred to as the subcarrier.
This second-order phase modulation,
or phase-modulated phase modulation,
will suppress ranging tone sidebands around the LISA
science fringe signal.

The current baseline plan for ranging tone modulation is
to phase modulate the USO signal before injection into the fiber modulator.
The modulation depth of the ranging tones will be half that
of the USO.  The frequency selection of the ranging tones
will be on the order of 100\,kHz.  This selection
yields a distance ambiguity of roughly 3\,km.
Varying this tone during acquisition, or using multiple tones
can reduce this ambiguity. \cite{refarch}

\subsection{Digital Data Transmission}

Complicating the situation further,
one operating mode for the LISA mission is expected to be that
digitally recorded phase data from each sciencecraft are sent to one selected
sciencecraft for data preprocessing.
This preprocessed data is then downloaded to Earth from the master sciencecraft.
Again, to prevent contamination of the LISA science data,
the digital data transmission between LISA sciencecraft will be phase modulated
onto the USO sideband.  Currently it is thought that digital data modulation
will happen in a similar way to the ranging tones described above.
The frequency of modulation most likely will be in the MHz range to
facilitate high data rates and good frequency separation from other signals.
One difference between the digital data modulation
and ranging tones or the clock tone is that the data modulation sideband
cannot be beat against another data sideband if any usable
information is to be extracted.  Instead, the digital data sideband
on the incoming beam will be beat against the USO subcarrier on the
local beam.  We describe this process in more detail in \S\ref{s:mod}.
Alternatively, both the digital phase data and the ranging tones might be
modulated onto the subcarrier using pseudo-random codes and $180^\circ$
phase-shift keying \cite{refarch}.

\subsection{Modulation Summary}

The LISA laser beam will be phase modulated with a phase modulated
subcarrier at gigahertz frequencies derived from the USO.
The subcarrier is modulated with ranging tones in the hundreds
of kilohertz and possibly data in the megahertz.
After traversing the five million kilometers between sciencecraft
the laser will have been Doppler shifted.  The Doppler shift ranges
in frequency up to 20\,MHz and changes at a rate up to about 1\,Hz/s.
Beatnotes between the incoming laser beam and the local laser beam will
contain the following signals to be measured:
the science signal from which the gravitational waves will be extracted
is the result of beating the two laser carrier frequencies, 
the USO signal is the result of beating the two USO subcarrier frequencies, 
the ranging signals are the result of beating different ranging sidetones,
the digital data signal is the result of beating the digital data
sideband on one laser with the USO subcarrier on another laser.

\bigskip
In this paper we present results from our table-top interferometer
with data modulation present on the laser beam.  In particular we
demonstrate the ability to transmit a clock tone with a ranging sidetone
and audio data on the laser beam from the distant sciencecraft and receive
that transmission.

\section{Laser Communication Design and Implementation}

In \cite{ifo} we presented a test-bed interferometer which produces
LISA-like fringes.  To investigate the effects of modulating information
on the LISA laser beam, we need only make slight modifications to our 
test-bed interferometer.  In particular, we add optical elements to modulate
data onto the laser beam and electronic components to read out the data.

As described in \S\ref{s:uso}, 
the baseline plan for LISA is to modulate the laser beams
using electro-optic modulators (EOMs).  
Our table-top interferometer simulates
one end of one arm of the LISA interferometer.
As described in \cite{jennrich}, we make use of a modified
Mach-Zehnder design which utilizes polarized light.
One arm of our interferometer is attenuated to 100\,pW
by using neutral density filters (NDFs).  
This ``dim beam'' represents the incoming light from
one of the far sciencecraft.  
The ``bright beam'' represents light from the
laser beam on the local sciencecraft.  The power 
of the bright beam is very nearly 0.5\,mW.  

\begin{sidewaysfigure}
\includegraphics[width=9.0in, angle=0]{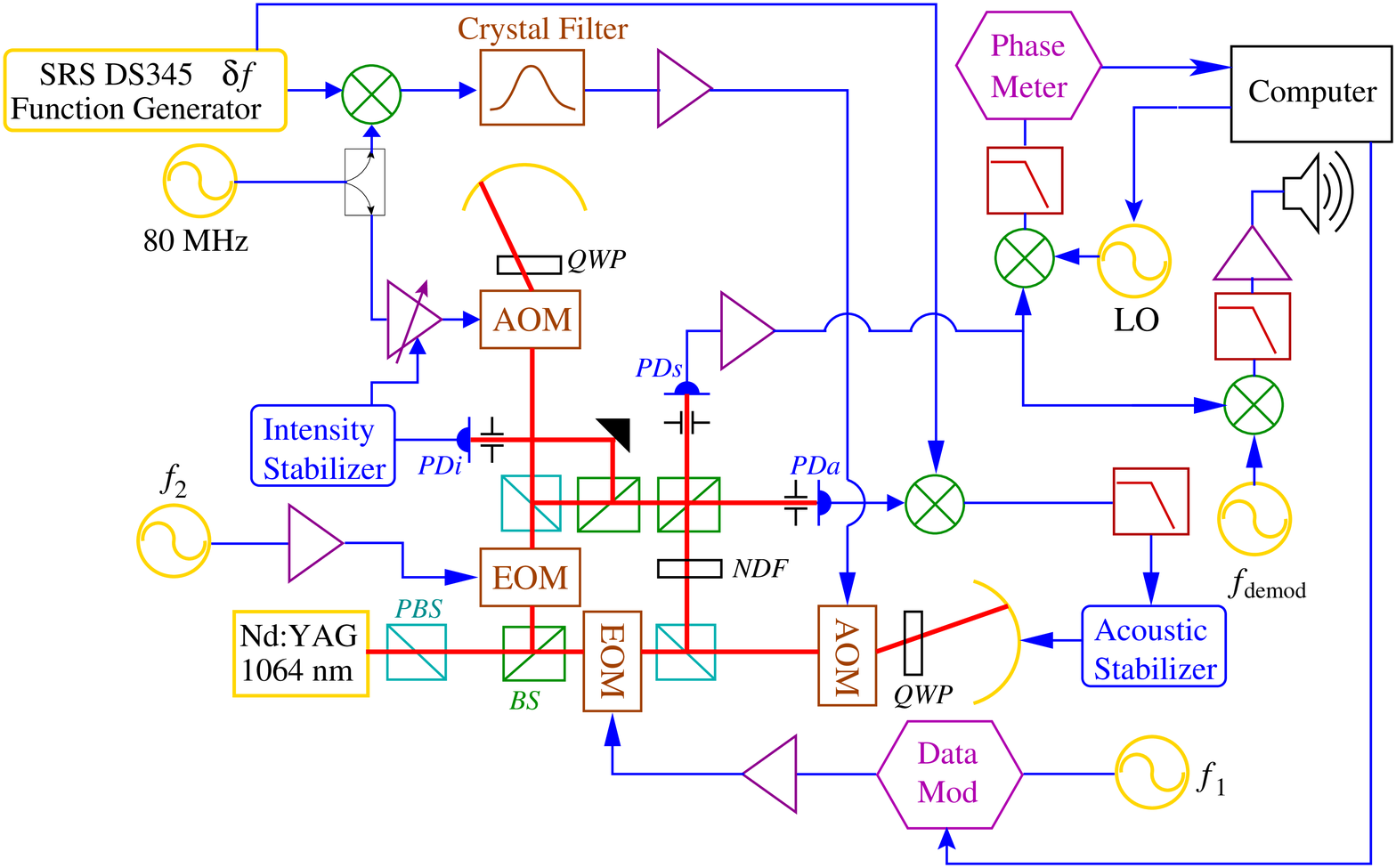}
\caption[Laser-Comm Interferometer Layout]{\label{fig:ifmcomm}
A schematic of our table-top fringe generator 
with laser communication components.
The electro-optic modulators (labeled ``EOM'') are placed before
the polarizing beamsplitters (PBSs) in each arm of the interferometer.
Data from the computer is modulated onto $f_1$
which in turn is phase modulated onto the laser
beam as described in \S\ref{s:mod}.  
The ``readout'' electronics consist of a mixer, a low-pass filter,
an amplifier, and an audio speaker.  The demodulation scheme is described
in detail in \S\ref{s:demod}.
}\end{sidewaysfigure}

Our experimental setup is shown in Figure~\ref{fig:ifmcomm}.
In order not to double pass our EOMs we place them before the 
polarizing beamsplitters (PBSs) in each arm.
For the dim arm, this placement simulates the 
LISA situation in which the laser beam is modulated
before attenuation by the pathlength distance between sciencecraft.

\subsection{Implementation of Data Modulation\label{s:mod}}

Our $5 \times 5 \times 20$\,mm birefringent crystals are made of 
lithium niobate (LiNbO$_3$).
We drive each EO crystal with
synthesized function generators.
The maximum frequency of these devices is 30\,MHz.  The maximum signal
level of these devices is 10\,volt-peak.  
For our setup, the resulting modulation depth with 
maximum signal level is about 3\%.  
The modulation depth planned for LISA is 10\% \cite{sts}.

Each frequency synthesizer provides us with 
the subcarrier frequency for data transmission.  
These signals are modulated with ranging tones and digital data and
then sent to the EOM, as shown in Figure~\ref{fig:ifmcomm}.
The frequencies $f_1$ and $f_2$ represent the USO
derived signals used in LISA.
In LISA the sideband-sideband beatnote 
will contain the relative phase jitters
between the USOs on-board the two sciencecraft.  
In our setup this beatnote contains phase jitters
resulting from the relative electronic phase noise in the 
two frequency synthesizers.

Ideally we would simultaneously simulate both 
the ranging tone modulation
and the digital data stream modulation.  
Due to hardware limitations, and for simplicity,
we have made measurements with only one of the modulations present
at any time.

Our frequency synthesizers have a built-in phase modulation option.  
The ranging tone sidebands, $f_{m1}$ and $f_{m2}$, 
are chosen to be in the hundreds of kHz.
Their separation, $| f_{m1} - f_{m2} |$, is in
the several tens of kHz.  
The ranging sideband-sideband beatnote
will be used as a vernier
to determine meter-scale distance changes between
the sciencecraft.  The larger scale distance
determination between sciencecraft will be
performed by ground-based measurements.

For simplicity, 
when simulating digital data modulation we have opted
to use the built-in analog amplitude modulation
port of our frequency synthesizers
rather than use a third-party 
data modulator, such as a phase-shift keyer.
We utilize this port to stream data from our computer to the EOM.  
When transmitting data we do not have ranging tones present
on the EO frequencies, $f_1$ or $f_2$.
For the case of digital data modulation 
we do not utilize a sideband-sideband beatnote.
Instead we beat the digital data sideband on
one laser beam with the subcarrier sideband
on the other laser beam.  This produces a signal with
frequency given by, e.g., $\delta\!f + | f_1 - f_2 | + f_d$, 
where $\delta\!f$ is the difference frequency between
the carriers,
and $f_d$ is the digital data modulation frequency.  Ideally
we would chose $f_d$ to be in the MHz.
Since we are not using an external data modulator
our digital data modulation frequency is limited to ten kilohertz.
For timely feedback on fidelity while aligning the interferometer
we stream audio data from the computer onto the laser for our
``digital'' data transfer.  The speaker shown in Figure~\ref{fig:ifmcomm}
is used for ``readout'' of this signal.

\begin{figure}[t]\centering
\includegraphics[width=6in, angle=0]{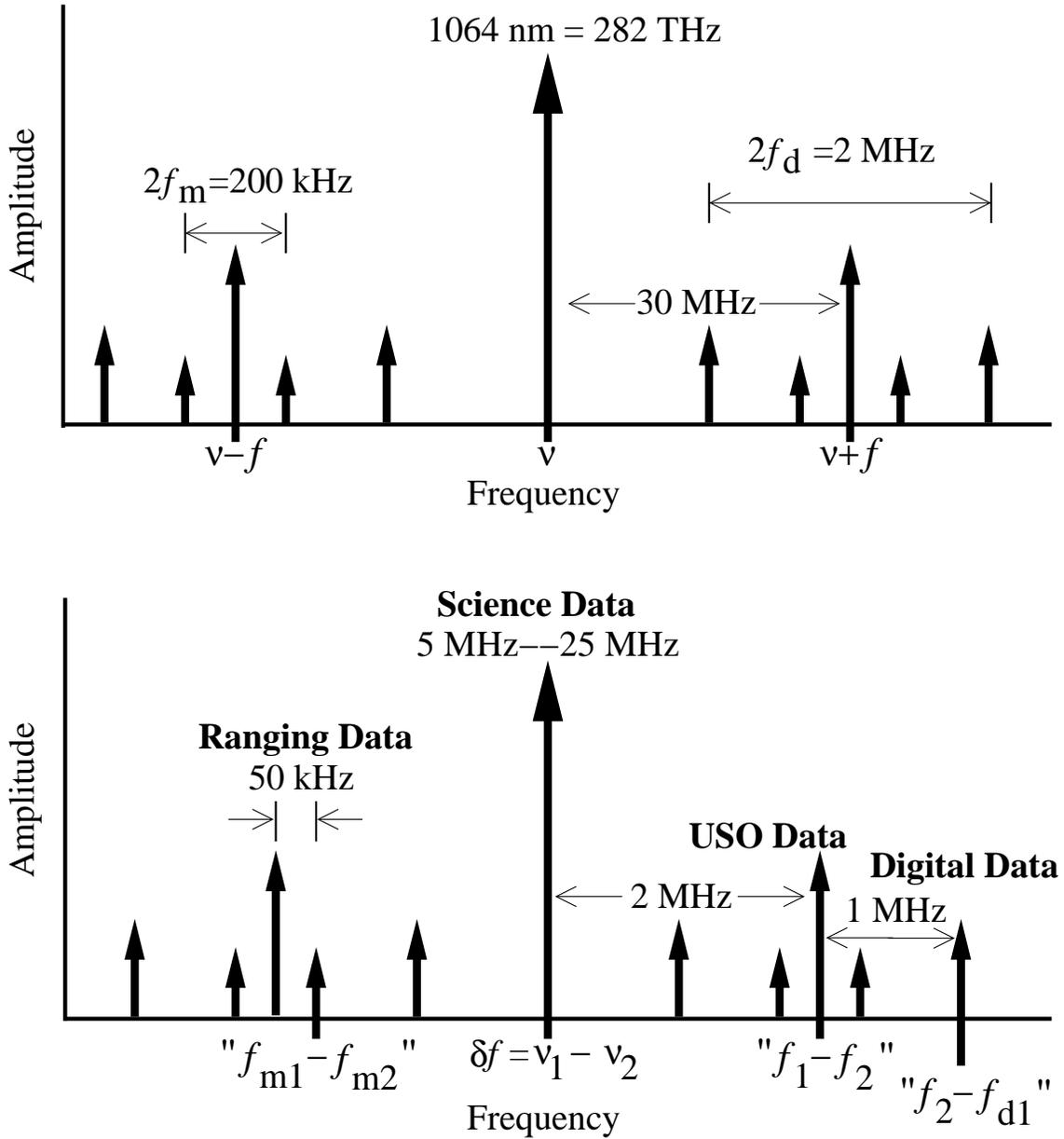}
\caption[Laser Intercommunication Demonstration Frequency Plan]{\label{fig:comm_freq}
Frequency plan for our laser intercommunication demonstration.  
The top chart represents the frequency plan for each arm of the interferometer
after the phase modulating EOM and before the frequency shifting AOM.  
The bottom chart shows the collection of beatnotes at the science
photoreceiver, PDs.  
}\end{figure}

Figure~\ref{fig:comm_freq} contains a summary of the frequencies used in 
our demonstration of sciencecraft intercommunication.  
The frequency $\nu$ is the laser frequency of our 1064~nm Nd:YAG.
The USO subcarrier and ranging tone frequencies, $f$ and $f_m$, 
are slightly different in the two arms.
The difference $| f_1 - f_2 |$ is chosen to be 2\,MHz,
and the difference $| f_{m1} - f_{m2} |$ is 50\,kHz.
Ideally our digital data sideband has a frequency of 1\,MHz,
as shown in the schematic.
In our setup the digital data signal, $f_d$, is audio
in nature and bound above in frequency by 10\,kHz 
from the carrier.

To prevent beatnotes from having negative frequencies after
the heterodyne process in the interferometer
we limit our carrier-carrier beatnote to be between 
5 and 25\,MHz (assuming a maximum Doppler shift due
to orbital motions of 20\,MHz).
The 5\,MHz offset can be accomplished in the control loop of
frequency locking the laser to a cavity or the LISA arms.

\subsection{Demodulation of Data\label{s:demod}}

\begin{figure}[t]\centering
\includegraphics[width=6in, angle=0]{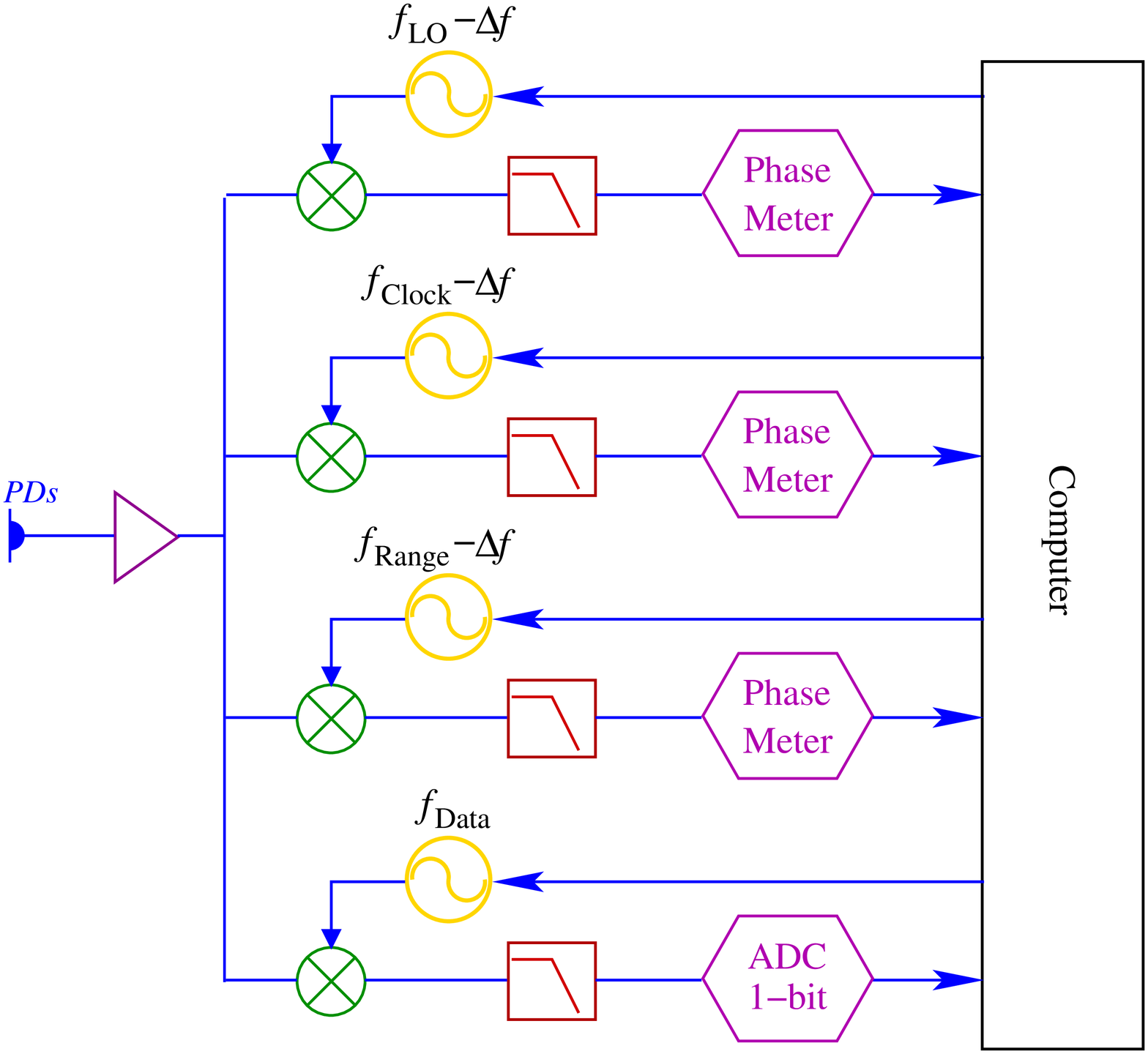}
\caption[Demodulation Plan]{\label{fig:demod}
Our proposed demodulation plan for use with the zero-crossing
phasemeter.  The intermediate frequency $\Delta\!f$ is chosen
appropriately for our phasemeter to be on the order of ten kilohertz.
Feedback from the computer adjusts the oscillators for the
slow Doppler frequency sweep due to orbital motions of the sciencecraft.
}\end{figure}

The electrical signals at the output of the photoreceiver
are shown in the bottom chart of Figure~\ref{fig:comm_freq}.
The demodulation scheme is depicted in Figure~\ref{fig:demod}.
The electrical signal from the photoreceiver is sent to
a distribution amplifier with four outputs.
Each output from the distribution amplifier is
the radio frequency (RF) input to a mixer.
The local oscillator (LO) input to each mixer is controlled 
by the computer so that the intermediate frequencies (IFs)
are suitable for our zero-crossing phasemeters 
(see \cite{ifo, pollack, jennrich}
for discussions of our zero-crossing phasemeter).

The first output from the distribution amplifier is
used for Doppler tracking as well as science extraction.
The carrier-carrier beatnote frequency is $\delta\!f = | \nu_1 - \nu_2 |$,
which would range from 5\,MHz to 25\,MHz due to the Doppler.
As explained above, the 5\,MHz offset is inserted
to prevent beatnotes from crossing DC.
The frequency $f_\mathrm{LO} \approx \delta\!f$ will approximate this
difference frequency.  Slow feedback from the computer
will keep the intermediate frequency from the first mixer
close to $\Delta\!f$, a frequency suitably chosen for our phasemeter.
The requirements on the tightness of
this feedback were discussed in Section 3 of \cite{ifo}.
Since all of the signals will sweep in frequency at essentially the
same rate, the sweep rate derived from the feedback of
$f_\mathrm{LO}$ will be used for the other demodulation
frequencies.

The second output from the distribution amplifier is
used in the extraction of the USO signal.  The USO data
is located in frequency space at $\delta\!f \pm | f_1 - f_2 |$.
The frequency $f_\mathrm{Clock}$ approximates this frequency:
$f_\mathrm{Clock} = f_\mathrm{LO} + 2\,\mathrm{MHz}$.  
Slow feedback from the computer will fix the IF
of the second mixer at $\Delta\!f$, suitable for our phasemeter.

The third output from the distribution amplifier is used
for extraction of ranging information.  In our frequency
plan, Figure~\ref{fig:comm_freq}, the ranging tone sideband-sideband
beatnote is at 50\,kHz.  This allows us to set the demodulation
frequency $f_\mathrm{Range} - \Delta\!f = f_\mathrm{Clock}$.
In this way the IF from the third mixer will be 50\,kHz.  Although
higher than our usual IF of 10\,kHz, 50\,kHz is an acceptable frequency
for our phasemeter.  As discussed in \cite{ifo}, our phasemeter can
accommodate IFs as high as 100\,kHz without being dominated
by timing noise errors.

Our proposal for the digital data signal is to use phase-shift keying.
The demodulation process would proceed as follows.
The data demodulation frequency approximates the expected
frequency of the digital data sideband beatnote:
$f_\mathrm{Data} = f_\mathrm{Clock} + 1\,\mathrm{MHz}$.
Instead of mixing the signal from PDs with $f_\mathrm{Data} - \Delta\!f$,
which would result in a phase-shift keyed $\Delta\!f$ signal,
we propose to mix the PDs signal with $f_\mathrm{Data}$.  
This results in a TTL-like signal jumping from high to low based
on the phase-shifts between the LO and the signal from PDs.  
This TTL-like signal can then be sampled with a 1-bit ADC.

As mentioned in \S\ref{s:mod}, for simplicity
we are modulating audio data onto our subcarrier
rather than a sub-subcarrier containing phase-shift keyed
digital data.
When audio data is present there are no ranging tones on
the subcarriers.  In this setup the subcarrier for the dim
arm of the interferometer is amplitude modulated
with audio data.  The demodulation scheme used
for this setup is that shown in the schematic of Figure~\ref{fig:ifmcomm}.
The demodulation frequency used for the audio data demodulation
is set to the USO frequency, $f_\mathrm{demod} = f_\mathrm{Clock}$,
just as in the case for the ranging data demodulation.
The demodulated output is amplified
and sent to a speaker rather than the phasemeter. 
When the speaker is turned on, the streaming
audio data sent by the computer is heard quite readily.

An alternative demodulation scheme 
would be to mix-out the signal in successive steps.
This scheme is similar to that which most likely would
be used with a digital phasemeter.  
At each step the signal is split into two paths.  
One path demodulates the signal to an IF suitable for the phasemeter,
and the other path demodulates the signal to ``DC''.
The first demodulation step is to take
out the carrier-carrier difference frequency of $\delta\!f$.
The Doppler is included in this demodulation step.
The next largest frequency component is the subcarrier-subcarrier
difference frequency.  After this demodulation the ranging tone
sideband-sideband beatnote and the digital data sideband remain.
The disadvantages with this demodulation scheme for an
analog phasemeter is the increased number of mixers
and frequency sources.  For a digital phasemeter,
these disadvantages are not present since the demodulation
process occurs in software.

\section{Data Comm Results}

We present data taken from our modified table-top interferometer 
with the zero-crossing phasemeter, presented in \cite{ifo},
of the science fringe signal, the USO signal, and the ranging signal.
In our setup the digital data signal was an analog audio
stream which was heard on a speaker as described above.
In all cases the dim arm of the interferometer is 100\,pW
and the bright arm of the interferometer is 0.5\,mW, as explained
in \cite{ifo}. 

To demonstrate that the EOM phase modulation
does not adversely affect the LISA science results we take
science fringe data at a collection of Doppler offset frequencies, $\delta\!f$, 
ranging from 50\,kHz to 20\,MHz, similar to the data presented in \cite{ifo}.
In all cases the Doppler frequency sweeps at a rate of 1\,Hz/s.  
In \cite{ifo} we demonstrated that our phasemeter is 
capable of handling sweep rates as fast as 1\,kHz/s.
The phase noise that we present is computed by subtracting
the known phase of our signal generator from the
reconstruction of the phase from our phasemeter (see \cite{pollack}).
From this set of residuals we compute an amplitude spectral density,
the phase noise, as a function of frequency from the signal frequency.

\subsection{Timing a LISA-like Fringe with 
Data Modulation Present\label{s:lisalike}}

The data presented in Figure~\ref{fig:specdata} were taken
while streaming audio was sent from the computer to the EO crystal
in the dim arm of the interferometer.   The signal to the EO crystal
in the bright arm of the interferometer was unmodulated.
This creates a data modulated signal on the fringe 2\,MHz
from the fringe frequency, which is the carrier-carrier beatnote.
The low-pass filter in our data demodulation scheme (see Figure~\ref{fig:demod})
removes the unwanted signals, such as the audio data and the 
subcarrier-subcarrier beatnotes,
and isolates the much larger amplitude science fringe frequency.

Each of the spectra in Figure~\ref{fig:specdata}
is the average of four 4.5-hour long data sets.  
For this data we instructed the LO to keep the IF between 1 and 2\,kHz.
Since the subcarriers are small in amplitude, and separated
in frequency by 2\,MHz,
the subcarrier frequency has little effect on the science results.  

There is some residual noise at approximately 3\,mHz.
Instructing the frequency synthesizer to output a constant
frequency removes this noise bump.  
Figure~\ref{fig:speccompare} shows data with swept and constant
frequency signals.
This increase in phase noise
most likely is caused by small changes in the phase delay
through the crystal filter used in the frequency
generation technique explained in \cite{ifo}.  
The phase delay through the crystal filter is frequency dependent.
This small change in phase appears to be
the source of the increased phase noise around 30\,mHz,
although why it peaks up near 30\,mHz is not clear.
This noise source was not
apparent in the data of \cite{ifo} because acoustic
noise was the dominating noise source below 100\,mHz.
The reduction of acoustic noise has been
mediated in this data by enclosing the interferometer in a
sound-insulating box.

\begin{idlplot}{p}{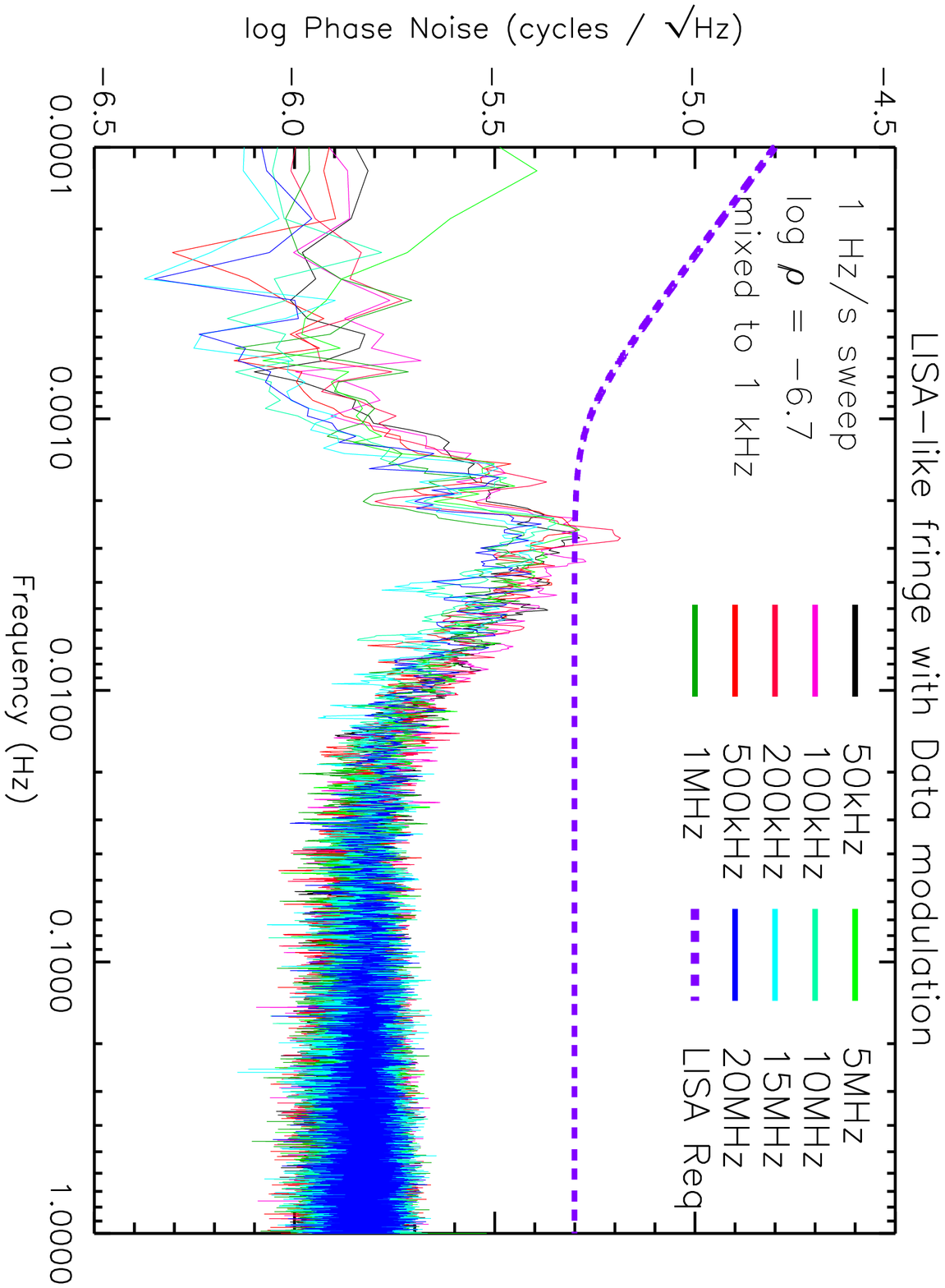}
\caption[Phase Noise of LISA-like Fringes with Data Modulation Present]
{\label{fig:specdata}
Phase noise of LISA-like fringes with a data modulation present.
The phase noise appears independent of baseband frequency
over the range examined.
See Figure~\ref{fig:speccompare} for comparision of phase
noise levels with and without data modulations.
}\end{idlplot}
\begin{idlplot}{p}{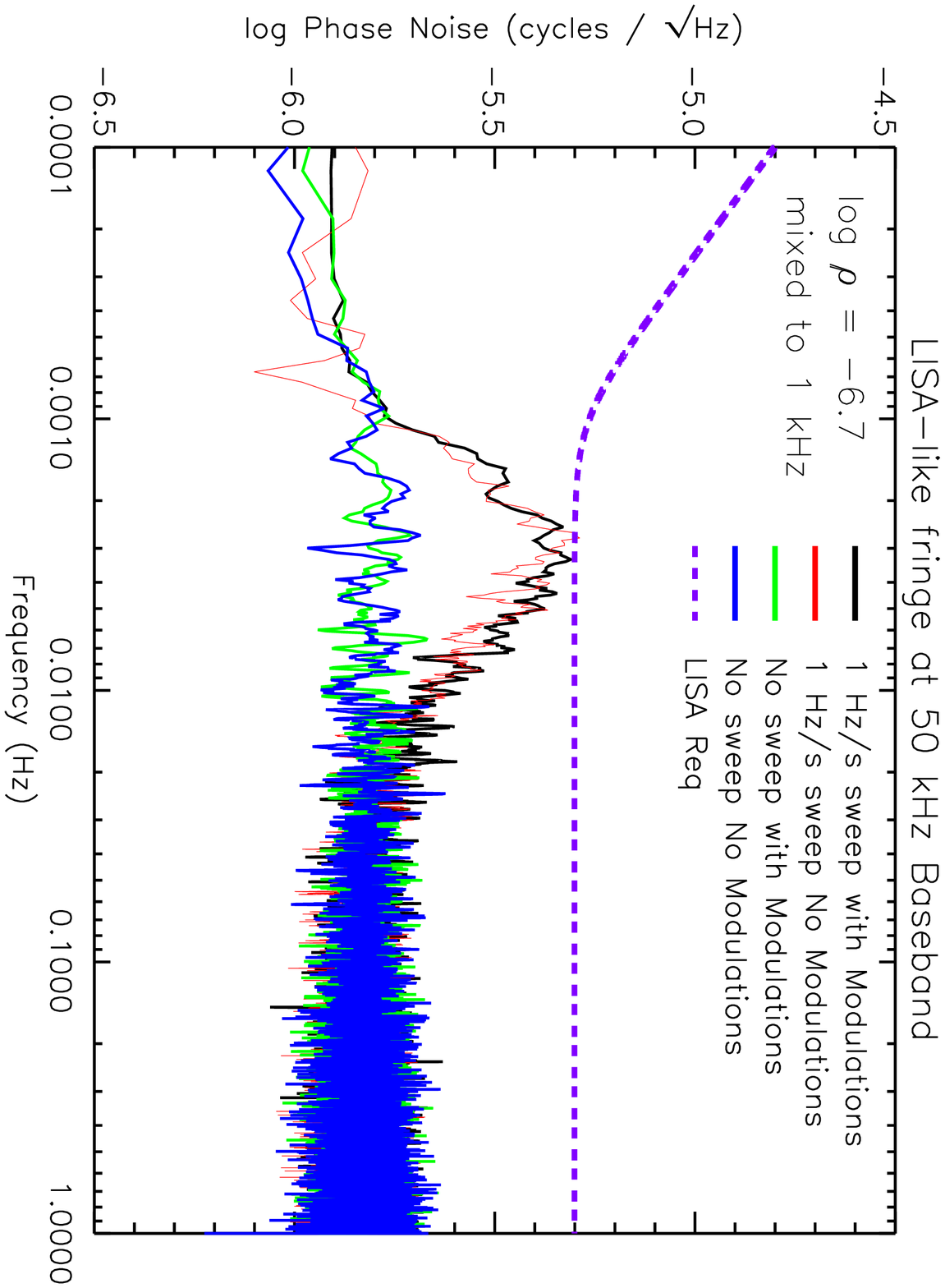}
\caption[Phase Noise of LISA-like Fringes at 50\,kHz Baseband]
{\label{fig:speccompare}
Comparision of phase noise of LISA-like fringes with and
without data modulations present.
The noise bump at 30\,mHz is present only for the
data where the interferometric fringe is sweeping
in frequency and is due to the frequency generation
technique as explained in the text.
The addition of data modulations does not hinder the science performance
of our interferometer.
}\end{idlplot}

\subsection{Timing the USO Subcarrier-Subcarrier Beatnote}

It is important not only to preserve the phase integrity of the science fringe
when data modulation is present, but also to preserve the data modulation
integrity itself.  In particular, as discussed at the beginning of this article,
the subcarrier will be a signal derived from the USO.
In the data presented in the previous section we amplitude
modulated the subcarrier in the dim arm with audio data.
This audio data was modulated onto the laser beam in our interferometer,
and we listened to a speaker to verify fidelity.

To determine the phase integrity of the subcarrier frequency which
represents the information transfer of an ultra-stable clock, we 
do not modulate on streaming audio.  
Audio frequencies range from 20\,Hz to 20\,kHz.  
If we wish to time the subcarrier-subcarrier beatnote by mixing it,
e.g., from $\delta\!f + 2\,\mathrm{MHz}$,
to a suitable IF for our phasemeter, e.g., $\Delta\!f = 10\,\mathrm{kHz}$, 
then our signal will be contaminated with irremovable audio data.

Modulating a sub-subcarrier onto the subcarrier at a much
higher frequency, e.g., 1\,MHz as shown in the frequency plan of 
Figure~\ref{fig:comm_freq}, 
would make filtering the audio data far more feasible.
Our frequency synthesizer has a maximum analog modulation frequency of 10\,kHz.
Using an external modulator, such as a phase-shift keyer with an
additional frequency source, 
would allow us to have a higher frequency modulation rate.

Instead of modulating audio data onto the subcarrier in the dim arm
while leaving the subcarrier in the bright arm unmodulated,
we phase modulate a 100\,kHz ranging tone onto one subcarrier
and a 150\,kHz ranging tone onto the other.
The sidebands present on the subcarrier-subcarrier beatnote
are the two ranging tones individually as well as the ranging tone
sideband-sideband beatnote.  Since the ranging tones
are in the hundreds of kilohertz and the ranging tone
beatnote is 50\,kHz, we use an IF of $\Delta\!f = 1$\,kHz 
with strong filtering to remove the unwanted signals.

\begin{idlplot}{t}{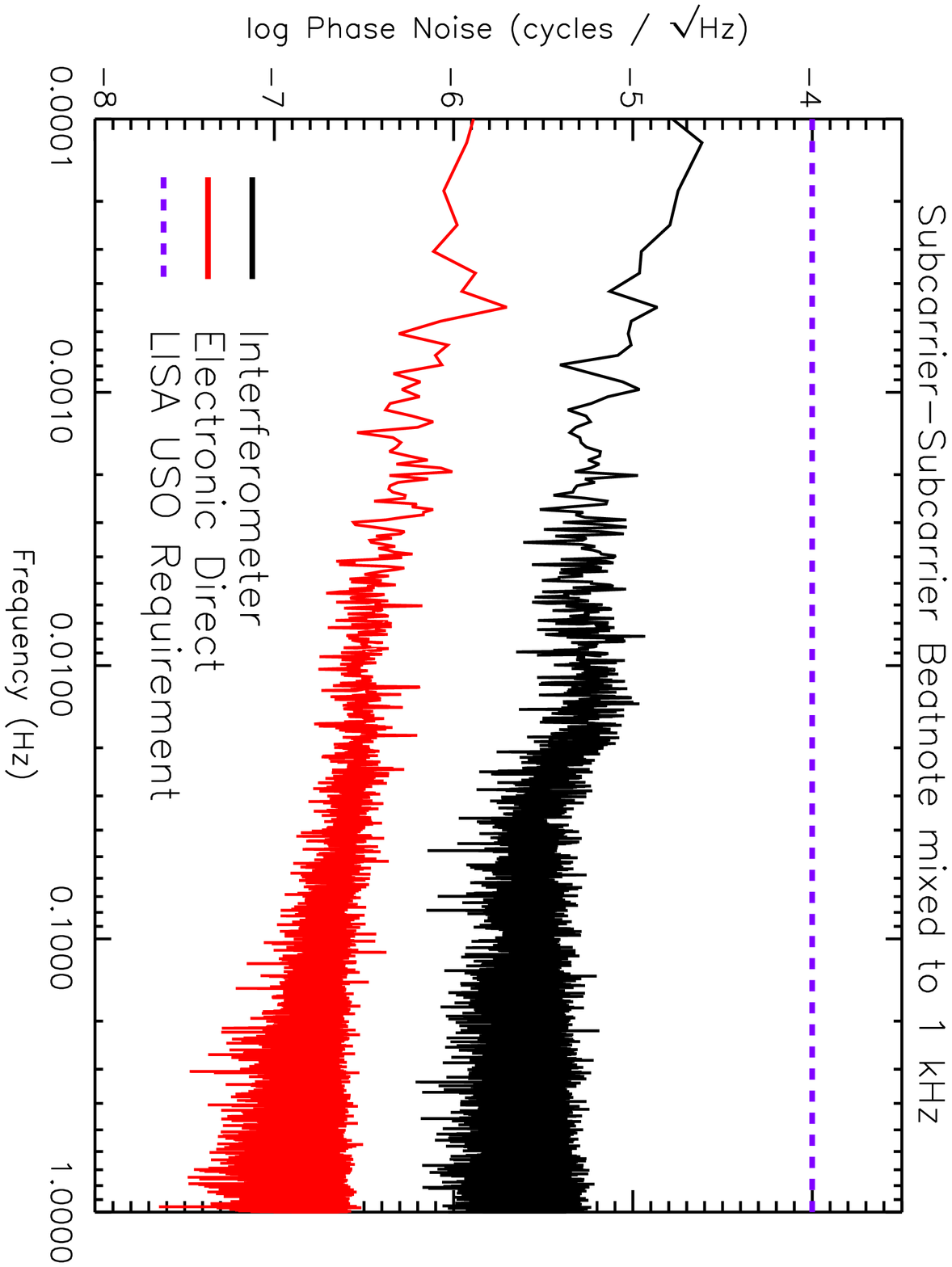}
\caption[Phase Noise of the USO Subcarrier Beatnote]{\label{fig:carrier}
Phase noise of the USO subcarrier-subcarrier beatnote.  The
ranging tone sideband-sideband beatnote is at 51\,kHz when
the USO beatnote is mixed to 1\,kHz.  
Strong filtering after mixing effectively removes
the ranging tone from this measurement.
The LISA USO requirement is set by TDI \cite{imsreq}.
}\end{idlplot}

Recall from \S\ref{s:uso} that the phase noise requirement
on the USO signal is less stringent than that on the fringe signal, 
the requirement is $10^{-4}\,\cph$ at 1\,mHz.
Figure~\ref{fig:carrier} contains the phase noise 
spectrum of the subcarrier-subcarrier beatnote 
signal passing through the interferometer as 
well as the phase noise spectrum of the subcarrier-subcarrier 
signal sent directly to the demixer, without passing
through the interferometer.  The latter case signal
was generated by mixing the two subcarriers together, sending
them through a low-pass filter, producing a $| f_1 - f_2 |$ signal, 
and then a demixer to a suitable IF, $\Delta\!f = 1$\,kHz.
The noise level of the subcarrier beatnote
is higher than that of the fringes presented in Figure~\ref{fig:specdata}
but still below the LISA USO requirement.
The increase in phase noise is due to the smaller
signal amplitude of the subcarrier beatnote on the output of the
photoreceiver as well as the fact that our acoustical
stabilization loop attempts to fix the phase of the fringe signal
but does not stabilize other signals \cite{ifo}.

\subsection{Timing the Ranging Tone Sideband-Sideband Beatnote}

Ranging tones may be modulated onto the subcarrier and used to assist
in the determination of the armlengths of the LISA interferometer.
The ranging tones on LISA will be in the hundreds of kilohertz,
and the modulation depths will be about 5\% of the total laser beam.
This is a 50\% modulation on the subcarrier.  We can demonstrate
this process by modulating tones onto our subcarriers in much the
same way that is planned to be done on LISA.

\begin{idlplot}{t}{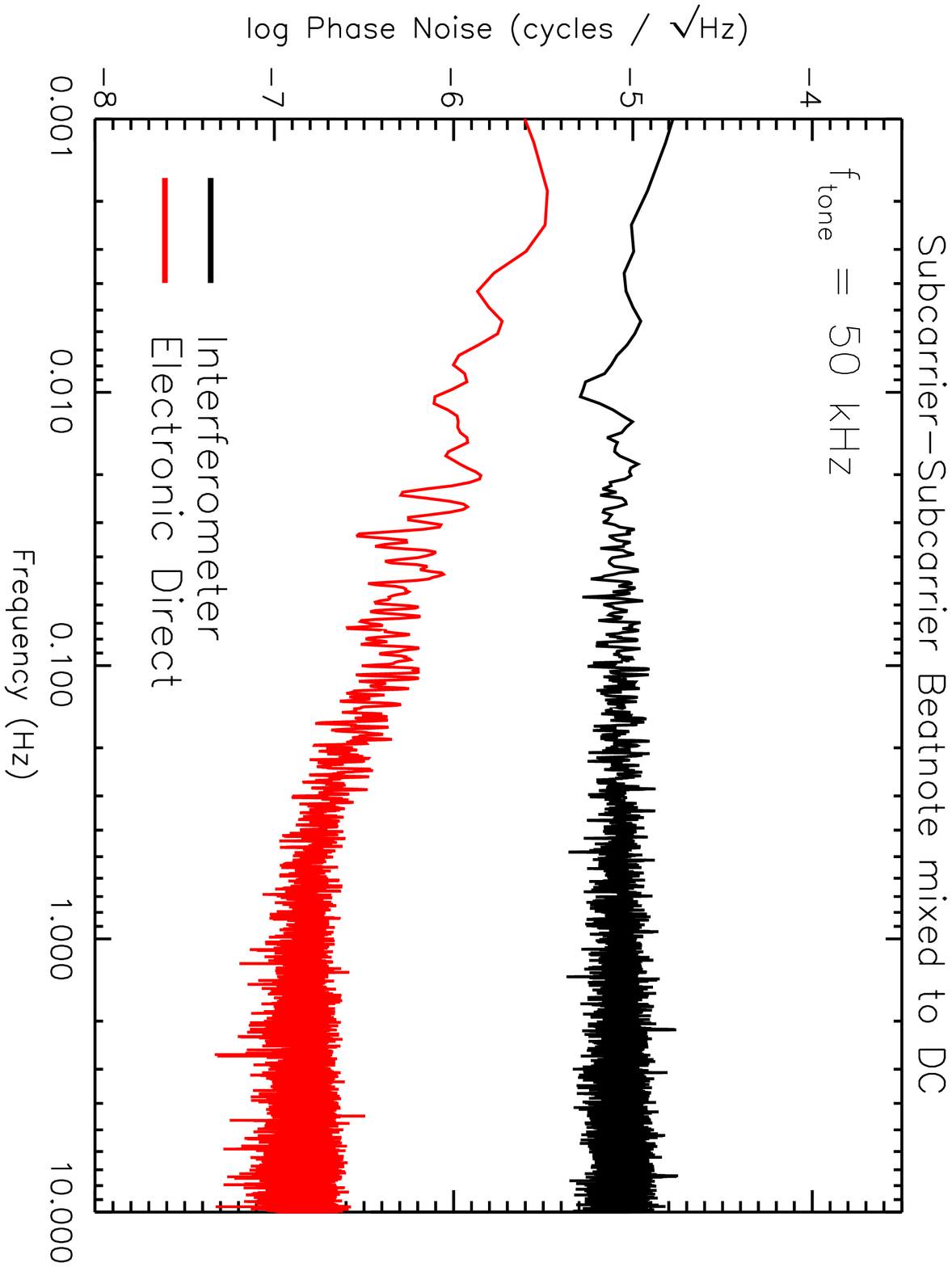}
\caption[Phase Noise of Monochromatic Tones]{\label{fig:tone}
100\,kHz and 150\,kHz tones are phase modulated onto 
onto the subcarriers.  The phase modulation depth was set to $90^\circ$.
The phase noise level is being dominated by voltage noise
in our photoreceiver circuit.
The assumed LISA ranging requirement of 20\,m is about 
$6.7 \times 10^{-3}\,\cph$,
well above the noise levels presented here.
}\end{idlplot}

As for the case of timing the USO subcarrier-subcarrier beatnote,
we do not modulate on streaming audio as this will confuse the
phase of our subcarrier and tone.  
We modulate one subcarrier with 100\,kHz and the other with 150\,kHz.
We demodulate the signal from the photoreceiver PDs with 
$f_\mathrm{Range} - \Delta\!f = f_\mathrm{Clock}$ as described
in \S\ref{s:demod}.  
The resulting signal frequency after filter is 50\,kHz.
Changes in the phase of this signal are a result from changes in 
the pathlength difference of our interferometer, akin to LISA.
Figure~\ref{fig:tone} contains phase noise data of this 50\,kHz
signal as well as the corresponding electronic signal created
by mixing the two subcarrier signals electronically.

There is no firm LISA requirement on the ranging tones.
However, it is assumed that a measurement sensitivity of $\Delta$L = 20\,m
would be a reasonable goal.  
At $f$ = 100\,kHz this is a phase measurement of $f\Delta$L/c = $6.7 \times 10^{-3}\,\cph$.  
Our data exceeds this measurement requirement.

\section{Summary\label{s:sum}}

We have modified our table-top interferometer presented in \cite{ifo}
with a laser communication system.  We have successfully modulated
hundred-kilohertz phase-modulated 28 and 30\,MHz signals onto our laser beam.
We have timed the phase of the ranging tone sideband-sideband beatnote
and the subcarrier-subcarrier beatnotes separately and have 
demonstrated that the phase noise levels are below the LISA requirements.  
For verification of data transfer
on the subcarrier we amplitude modulated the subcarrier in the dim arm with
analog audio data.  
During this exercise the subcarrier in the bright arm of the interferometer
was unmodulated.
This data was received, demodulated, and
the audio data was played back on a speaker. 

There are a number of differences between our setup and the one which
will be implemented in LISA.  
We have phase modulated each arm of our interferometer
with signals derived from phase locked frequency
synthesizers.  In the case of LISA the phase modulations
will be derived from independently running USOs.
The phase of the heterodyned beatnote is measured.
In the case of LISA this phase information contains
the relative phase fluctuations between the USOs
on-board the two sciencecraft, and is used in the TDI algorithm
to correct for laser frequency noise.
In our case this phase information contains
the relative jitter between our two phase locked
frequency synthesizers.  The phase noise level of the
USO signal we have measured is much smaller than that expected for
LISA.  Regardless of the level of the noise, the measurement
precision required by TDI is $10^{-4}\,\cph$.
We have measured the noise to a precision of $10\,\mcph$.

Another difference between our experiment and LISA is
the choice of frequencies.  The USO subcarrier frequency
will be in the gigahertz range on LISA, whereas we used 28 and 30\,MHz.
Using a much higher frequency further reduces the effects on the
science fringe.  
In addition, having a high frequency phase modulation, such as 2\,GHz,
reduces the effect of laser frequency noise contamination
of the subcarriers.
In our experimental setup, since we use one laser for the heterodyne
measurement, the laser frequency noise contribution to our measurement
is effectively absent.

An improvement to our experimental setup would be to better demonstrate
the full LISA laser communication plan described in \S\ref{s:mod}.
In particular we would use an external phase-shift keyer to
modulate sub-subcarrier signals at $f_{d1}$ and $f_{d2}$.  
These frequencies might be 1\,MHz and 3\,MHz (see Figure~\ref{fig:comm_freq}).
These sub-subcarriers would then be phase modulated onto the subcarriers
$f_1$ and $f_2$ before modulation onto the laser beams with the EOMs.
Demodulation of the science photoreceiver signal 
by $f_\mathrm{Data} = f_\mathrm{Clock} + 1\,\mathrm{MHz}$
or by $f_\mathrm{Data} = f_\mathrm{Clock} + 3\,\mathrm{MHz}$
would produce a TTL-like signal containing the phase-shift 
keyed data.

\ack

Support for this work has been provided under NGT5-50451 and S-73625-G.
We would like to thank Peter L. Bender, John Hall, and Jun Ye for their
generous insight to aspects of this project.

\end{document}